\documentstyle[aps,epsf]{revtex}

\begin{document}

\baselineskip 14pt

\title{Hadronic Three Jet Production at Next-to-Leading Order}
\author{William B. Kilgore}
\address{Brookhaven National Laboratory, Upton,
NY 11973-5000, USA}
\author{Walter T. Giele}
\address{Fermi National Accelerator Laboratory, P.O. Box 500, Batavia,
IL 60510, USA}

\maketitle

\begin{abstract}
We report preliminary results for a next-to-leading order event
generator for hadronic three jet production.  We demonstrate the
stability of the calculation and present preliminary results for the
jet transverse energy spectra.  This is the first calculation of three
jet production at this order to include all parton sub-processes.
\end{abstract}

\section{Introduction}
In this talk I will discuss recent work in constructing a
next-to-leading order event generator for hadronic three jet
production.  This is the first calculation of three jet production at
this order to include all parton sub-processes.  Previous
studies~\cite{Troc,KG} have only included the contributions of pure
gluon scattering contribution.  As my preliminary results show,
the generator is now working properly and is ready to perform
phenomenological studies.

\section{Motivation}
%\subsection{Why NLO?}
When interpreting experimental data, one would like to have some
understanding of the uncertainty associated with theoretical
expectations.  In QED and the weak interactions this is not a big
problem because the couplings are sufficiently weak that higher order
corrections are generally quite small.  In QCD, however, the coupling
is quite strong ($\alpha_s$ is still of order $1/8$ at the scale of
the $Z$ boson mass) and it is difficult to obtain a reliable estimate
of the theoretical uncertainty.  Typically, one characterizes
theoretical uncertainty by the dependence on the renormalization scale
$\mu$. Since we don't actually know how to choose $\mu$ or even a
range of $\mu$, the uncertainty associated with scale dependence is
somewhat arbitrary.  There seems no way around this other than to
calculate to higher order where the scale dependence is expected to be
smaller. 

However, one often obtains other improvements besides reduced scale
dependence by going to higher order.  Sometimes one finds the
next-to-leading order (NLO) correction to be very large.  A notorious
example is Higgs production at hadron colliders where the NLO
corrections to the leading order gluon fusion process are of the order
of $100\%$.  Such large corrections often come from opening up new
channels that are forbidden at leading order (LO), but one must still
be concerned about the question of perturbative convergence.
Regardless of the nominal scale dependence, one simply does not know
how to trust a calculation when the perturbative corrections are
large.

Even if the overall NLO correction is relatively small, there may be
regions of phase space, typically near the boundaries of the allowed
region for the LO process, where NLO corrections are large.  In these
regions, NLO calculations are effectively of leading order and suffer
from the large scale dependence associated with leading order.  It is
only in those regions of phase space where the NLO corrections are
well behaved (as determined by the ratio of the NLO to LO terms) that
one has confidence in the reliability of the calculation and can begin
to believe the uncertainty estimated from scale dependence and it is
only when one has a reliable estimate of the theoretical uncertainty
that comparisons to experiment are meaningful.

%\subsection{Why NLO Three Jets?}
A next-to-leading order three jet calculation will have many
phenomenological applications.  One of the most important will be to
perform a purely hadronic extraction of $\alpha_s$ via the ratio of
two-jet to three-jet production.  Because these processes have the
same production mechanisms, such an extraction should be relatively
free of parton distribution uncertainties.  Since hadron machines
produce events at all accessible energy scales, it should be possible
to measure the running of $\alpha_s$ and thereby the QCD $\beta$
function which depends upon the strongly interacting matter content
accessible at each scale.

This calculation will also be useful for studying jet algorithms.  One 
would like to have a flexible jet algorithm that makes it easy to
compare experimental results with theoretical calculations.  The
presence of as many as four final state partons permits more
complicated clustering conditions and tests the details of the
algorithms.  In our pure gluon study~\cite{KG}, we found that the
iterative cone algorithms commonly used at hadron colliders have an
intrinsic infrared sensitivity that precludes their direct
implementation in fixed order calculations.  With Run II at the
Tevatron fast approaching, it would be desirable to settle on an
algorithm suitable for both theory and experiment.

Other applications include the study of energy flow within jets and
background studies for new phenomena searches.  Finally, this entire
calculation is but a part of an eventual next-to-next-to-leading order
(NNLO) calculation of two-jet production.  That calculation is still a
long way off.  Not only are the two-loop virtual corrections unknown,
but even the higher-order contributions to real emission are unknown.
Still, the NNLO calculation will eventually be needed to compare to
the high statistics data that will be collected at the Tevatron and
the LHC. 

\section{Methods}
The NLO three jet calculation consists of two parts: two to three
parton processes at one-loop (the virtual terms) and two to four
parton processes (the real emission terms) at tree-level.  Both of
these contributions are infrared singular; only the sum of the two is
infrared finite and meaningful.  The virtual contributions are
infrared singular because of loop momenta going on-shell.  The virtual
singularities take the form of single and double poles in the
dimensional regulator $\epsilon$ multiplying the Born amplitude.  The
real emission contributions are singular when two partons become
collinear or when a gluon becomes very soft.  The
Kinoshita-Lee-Nauenberg (KLN) theorem~\cite{kln} guarantees that the
infrared singularities cancel for sufficiently inclusive processes
when the real and virtual contributions are combined.

The parton sub-processes involved are $gg\to ggg$~\cite{BDK1},
$\overline{q}q\to ggg$~\cite{BDK2},
$\overline{q}q\to\overline{Q}Qg$~\cite{KST}, and processes related to 
these by crossing symmetry, all computed to one-loop, and $gg\to
gggg$, $\overline{q}q\to gggg$, $\overline{q}q\to\overline{Q}Qgg$, and
$\overline{q}q\to\overline{Q}Q\overline{Q^\prime}Q^\prime$ and the
crossed processes computed at tree-level.  Quark-antiquark pairs
$Q\overline{Q}$ may or may not have the same flavor as the 
$q\overline{q}$ and $Q^\prime\overline{Q^\prime}$ pairs.
Previous NLO three jet calculations have worked in the approximation
of pure gluon scattering, using only the $gg\to ggg$ and $gg\to gggg$
processes. This is the first calculation to include all parton
sub-processes.

In order to implement the kinematic cuts necessary to compare a
calculation to experimental data one must compute the cross section
numerically.  Thus, it is not sufficient to know that the
singularities drop out in the end, we must find a way of canceling
them before we start the calculation.  The crucial issue in obtaining
and implementing the cancelation is
resolution.  The real emission process is infrared singular in
precisely those regions where the individual partons cannot all be
resolved (even in principle, ignoring the complication of
hadronization, showering, {\it etc.}) because of collinear overlap or
by becoming too soft to detect.  If we impose some resolution
criterion, we can split the real emission calculation into two parts,
the ``hard emission'' part in which all of the partons are well
resolved and the ``infrared'' part in which one or more partons are
unresolved.

The hard emission part is computed in the normal way by means of Monte
Carlo integration.  The infrared part is treated differently, making
use of the fact that matrix elements have well defined factorization
properties in both soft and collinear infrared limits.  In terms of
color-ordered helicity amplitudes, 
\begin{eqnarray}
{\cal M}_n(\ldots,1^{\lambda_1},2^{\lambda_2},\ldots)\ 
{\buildrel1\parallel2\over\longrightarrow}\ {\mathop{\rm
Split}\nolimits}_{-\lambda_c}(1^{\lambda_1},2^{\lambda_2}){\cal
M}_{n-1}(\ldots,c^{\lambda_c},\ldots) \nonumber\\
\\
{\cal M}_n(\ldots,1^{\lambda_1},s^{\lambda_s},2^{\lambda_2},\ldots)\
{\buildrel k_s\to0\over\longrightarrow}\ {\mathop{\rm Soft}\nolimits}
(1,s^{\lambda_s},2){\cal
M}_{n-1}(\ldots,1^{\lambda_1},2^{\lambda_2},\ldots), \nonumber
\end{eqnarray}
where ${\mathop{\rm Split}}$ and ${\mathop{\rm Soft}}$ are
universal functions depending only on the momenta, helicities and
particle types involved.  The ${\mathop{\rm Split}}$ functions are
in a sense the square roots of the Altarelli-Parisi splitting
functions.  In computing the infrared part, we replace the full
two-to-four parton matrix elements with their infrared factorized
limits.  We then integrate out the unresolved parton by integrating
(in dimensional regularization) the ${\mathop{\rm Split}}$ and
${\mathop{\rm Soft}}$ functions over the unresolved region of phase
space, resulting in single and double poles multiplying two-to-three
parton Born matrix elements.  These terms, as the KLN theorem says
they must, have exactly the right pole structure to cancel the
infrared poles of the virtual contribution.  By analytically combining
unresolved real emission with the virtual terms, we obtain a finite
contribution that can be integrated numerically.

Several different methods~\cite{BOO,KS,GG,GGK,CS,FKS,KG} of
implementing this infrared cancelation 
have successfully employed in various NLO calculations.  The method we
use is the ``subtraction improved'' phase space slicing
method~\cite{KG}.  The phase space slicing method~\cite{GG,GGK} uses a
resolution criterion $s_{min}$, which is a cut on the two parton
invariant masses,
\begin{equation}
s_{ij} = 2E_iE_j(1-\cos\theta_{ij}).
\end{equation}
If partons $i$ and $j$ have $s_{ij}>s_{min}$ they are said to
be resolved from one another.  (Which is not to say that a jet
clustering algorithm will not put them into the same jet.)  If
$s_{ij}<s_{min}$ partons $i$ and $j$ are said to be unresolvable.
One advantage of the $s_{min}$ criterion is that it simultaneously
regulates both soft ($E_i\to0$ or $E_j\to0$) and collinear
($\cos\theta_{ij}\to1$) emission.  In the rearrangement of terms, the
infrared region of phase space is where any two parton invariant mass
is less than $s_{min}$.  These regions are sliced out of the full
two-to-four body phase space, partially integrated and then added to
the two-to-three body integral.

Because the infrared integral is bounded by $s_{min}$, the
integrations over ${\mathop{\rm Split}}$ and ${\mathop{\rm Soft}}$
terms depend explicitly on $s_{min}$.  In fact, in the cancelation of
the virtual singularities, the $1/\epsilon$ terms are replaced by
$\ln{s_{min}}$ terms and the $1/\epsilon^2$ terms by $\ln^2{s_{min}}$
terms.  The hard real emission term is also $s_{min}$ dependent because
the boundary of the sliced out region depends on $s_{min}$.  Because
$s_{min}$ is an arbitrary parameter the sum of the virtual and real
emission terms must be $s_{min}$ independent.  Thus, we have
rearranged the calculation, trading a cancelation of infrared poles in
$\epsilon$ for a cancelation of logarithms of $s_{min}$.  This
provides an important cross check on our calculation.  If we can
demonstrate that our calculated cross section is $s_{min}$ independent
we can be confident that we have correctly implemented the infrared
cancelation. 

While the NLO cross section is formally independent of $s_{min}$,
there are several practical considerations to choosing the value
properly.  As $s_{min}$ becomes smaller, the infrared approximations
of the matrix elements becomes more accurate.  However, the overriding
concern is the numerical convergence of the calculation.  Two terms,
each diverging like $\ln^2s_{min}$ must be added with the logs
canceling.  As $s_{min}$ is made small, the logarithm becomes large
and the individual terms, real and virtual, become larger in
magnitude.  The sum however, the NLO cross section, is unchanged.
Thus, as $s_{min}$ becomes small, it becomes harder to engineer the
cancelation to the precision to which one would like to compute the
cross section.  Based on this consideration, we would like to make
$s_{min}$ as large as possible.  There is an absolute upper limit
imposed by the constraints of jet finding.  We cannot make $s_{min}$
so large that it begins to interfere with jet clustering, say, by
declaring unresolvable a pair of partons that a sensible jet
clustering algorithm would say are separated from one another.

Another problem with large values of $s_{min}$, alluded to before and
which actually sets in at a lower scale than jet clustering
interference, is that as $s_{min}$ is made larger the infrared
approximations used in the slicing region become less precise.  The
``subtraction improved'' part 
of our method involves the handling of the slicing region.  As
originally implemented, the infrared region was completely sliced out
of the two-to-four integration and the full two-to-four matrix element
was replaced by its soft or collinear limit.  A better approximation
is to leave the infrared regions in the ``hard'' phase space integral,
but to compute only the difference between the true and approximate
matrix elements in those regions.  In our gluonic three jet production
study~\cite{KG} we found that the subtraction improvement allowed us
to use substantially larger values of $s_{min}$ than would have been
possible with just phase space slicing.

\section{Preliminary Results}
The results presented below were computed for the following kinematic
configurations: the $\bar{p}p$ center of mass energy is $1800$ GeV; 
the leading jet is required to have at least $100$ GeV in transverse
energy, $E_T$, and there must be two additional jets with at least
$50$ GeV of transverse energy; all jets must lie in the pseudorapidity
range $-4.0 < \eta_J < 4.0$.  The CTEQ3M parton distribution
functions~\cite{CTEQ3M} and the EKS~\cite{EKS} jet clustering
algorithm (modified for three jet configurations as in
reference~\cite{KG}) were used. 

\begin{figure}[ht]	% in second brace, h=here, t=top, b=bottom	
\hbox to \hsize{\hfil 
   \vbox{\epsfxsize 6truein\epsfbox{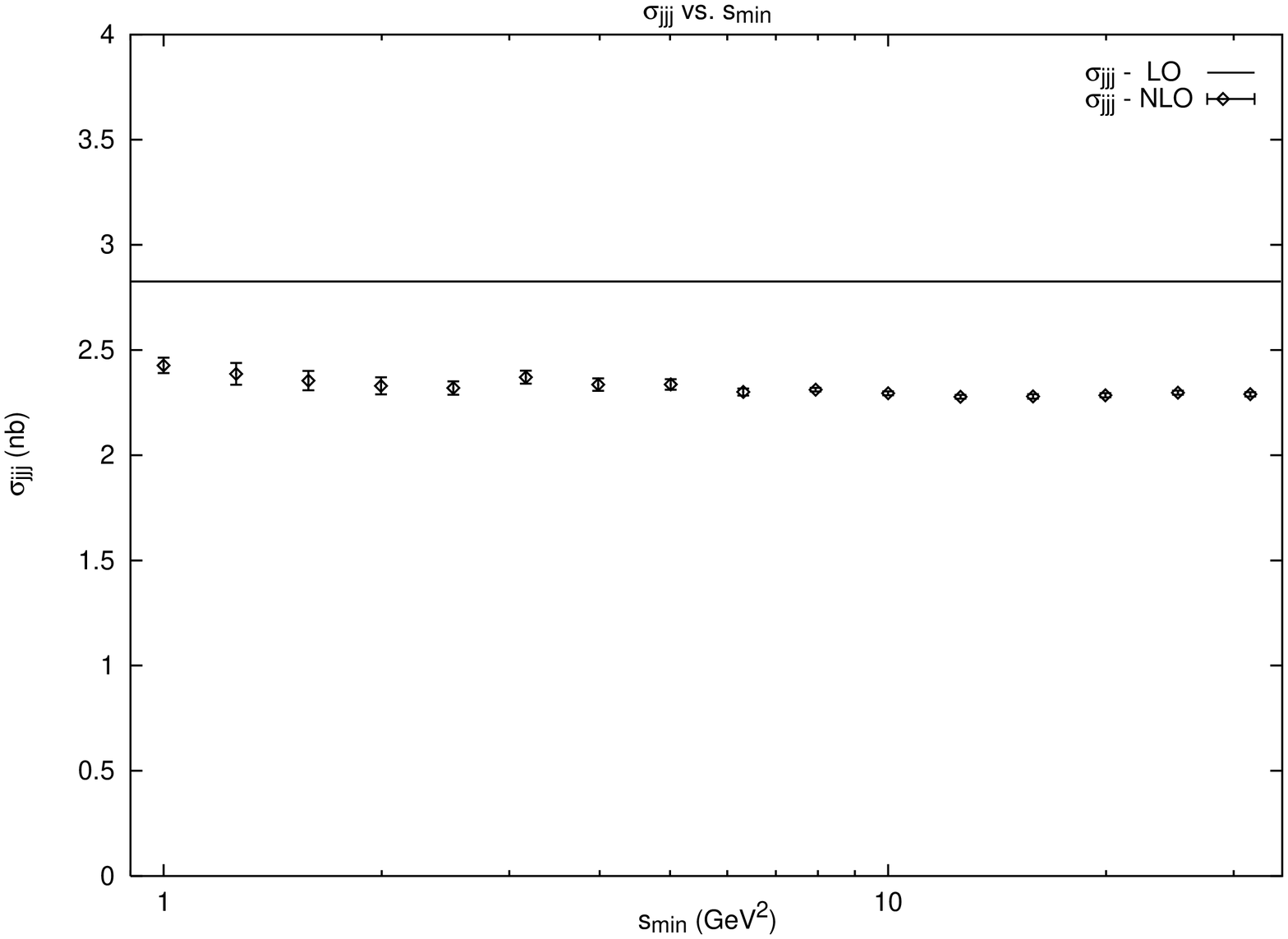}
         \epsfxsize 6truein\epsfbox{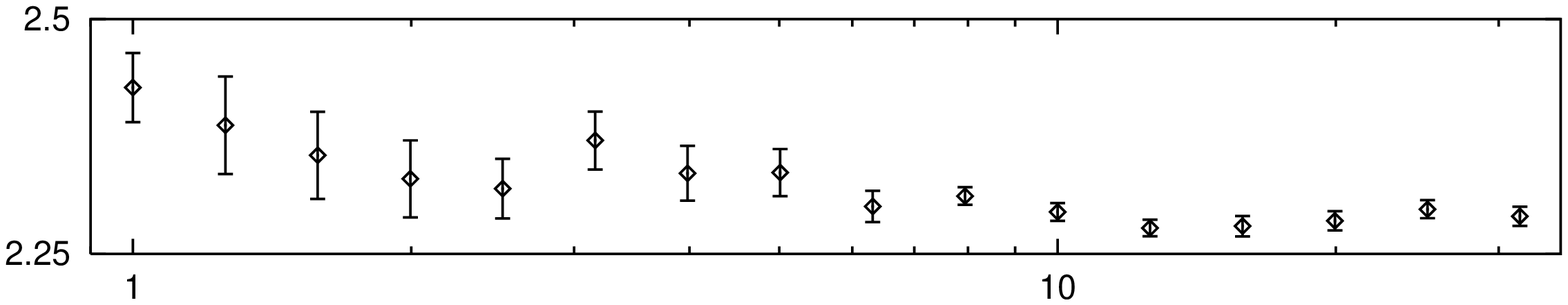}}\hfil}   
\caption[]{
\label{fig:sminplot}
\small Top: Next-to-leading order three jet cross section vs. $s_{min}$.
The leading order cross section is shown as a solid line.
Bottom: Expanded view showing relative statistical uncertainties.}
\end{figure}

Figure~\ref{fig:sminplot} shows the computed next-to-leading order
three jet cross section as a function of the resolution parameter
$s_{min}$.  Also shown is the leading order calculation.

We see that the NLO result is stable over a wide range of values of
$s_{min}$.  This stability indicates that we are correctly
implementing the infrared cancelation.  Further calculations at larger
values of $s_{min}$ are needed to actually determine the limit of the
region of stability.  In the lower plot, we see the statistical
uncertainty on each point. As $s_{min}$ becomes small, 
it becomes increasingly difficult to calculate $\sigma_{jjj}$ to the
desired precision.  For instance, at $s_{min}=10$ GeV${}^2$, the real
and virtual components are $16.763\pm0.008$ and $-14.468\pm0.005 (nb)$
respectively, while at $s_{min}=1$ GeV${}^2$, they are
$29.739\pm0.035$ and $-27.312\pm0.010$.  To 
obtain the same absolute uncertainty on the sum of these numbers, the
relative uncertainty on each of the components at $s_{min}=1$
GeV${}^2$ must be one half that required at  $s_{min}=10$ GeV${}^2$.
Since the statistical uncertainty scales like the square root of
number of points evaluated, it takes roughly four times as long to
obtain a precise calculation at $s_{min}=1$ GeV${}^2$ as it does at
$s_{min}=10$ GeV${}^2$. 

We also see that the size of the NLO correction is of order $15\%$.
This gives us confidence in the perturbative stability of the
calculation.  Together, these two observations indicate that we are
performing a reliable calculation of the three jet cross section.
Further tests using a variety of modern parton distribution functions,
values of $\alpha_s$, renormalization scales, {\it etc.\/} are needed
to obtain a clearer picture of the theoretical uncertainty associated
with the calculation.

\begin{figure}[ht]	% in second brace, h=here, t=top, b=bottom	
\hbox to \hsize{\hfil \epsfxsize 6.0 truein\epsfbox{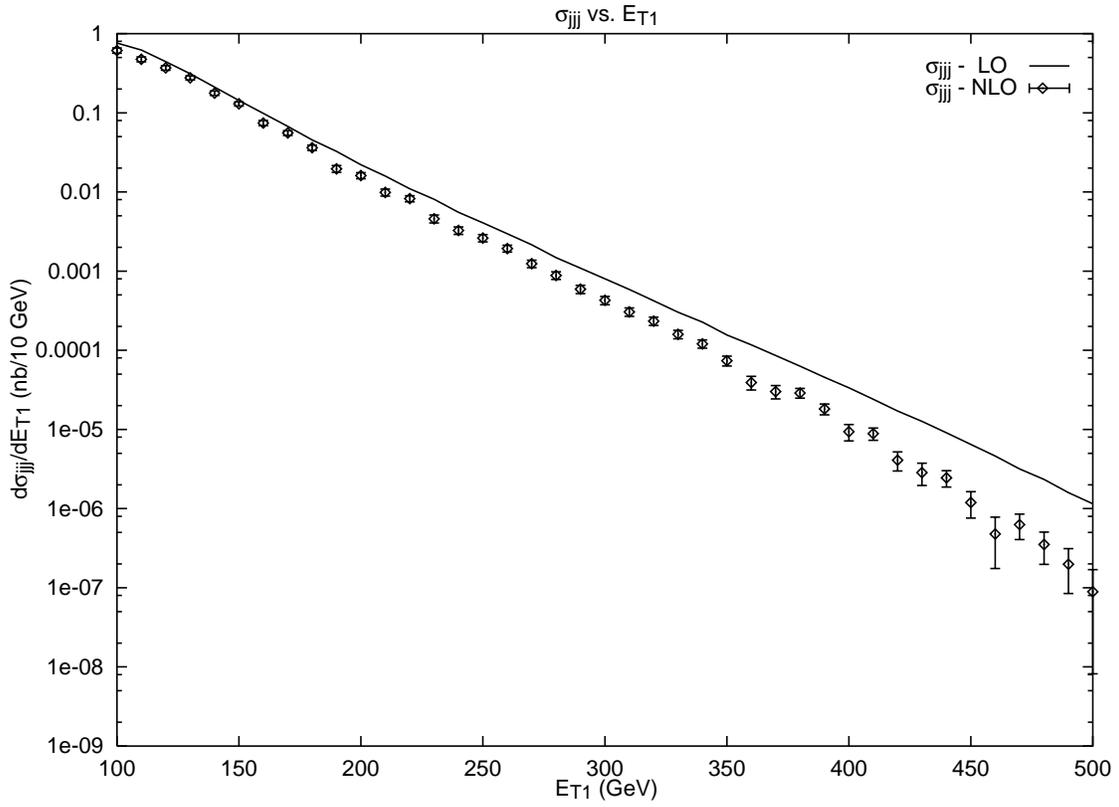}\hfil}   
\caption[]{
\label{fig:ET1plot}
\small Transverse energy spectrum of the leading jet.
The leading order result is shown as a solid line.}
\end{figure}

\begin{figure}[ht]	% in second brace, h=here, t=top, b=bottom	
\hbox to \hsize{\hfil \epsfxsize 6.0 truein\epsfbox{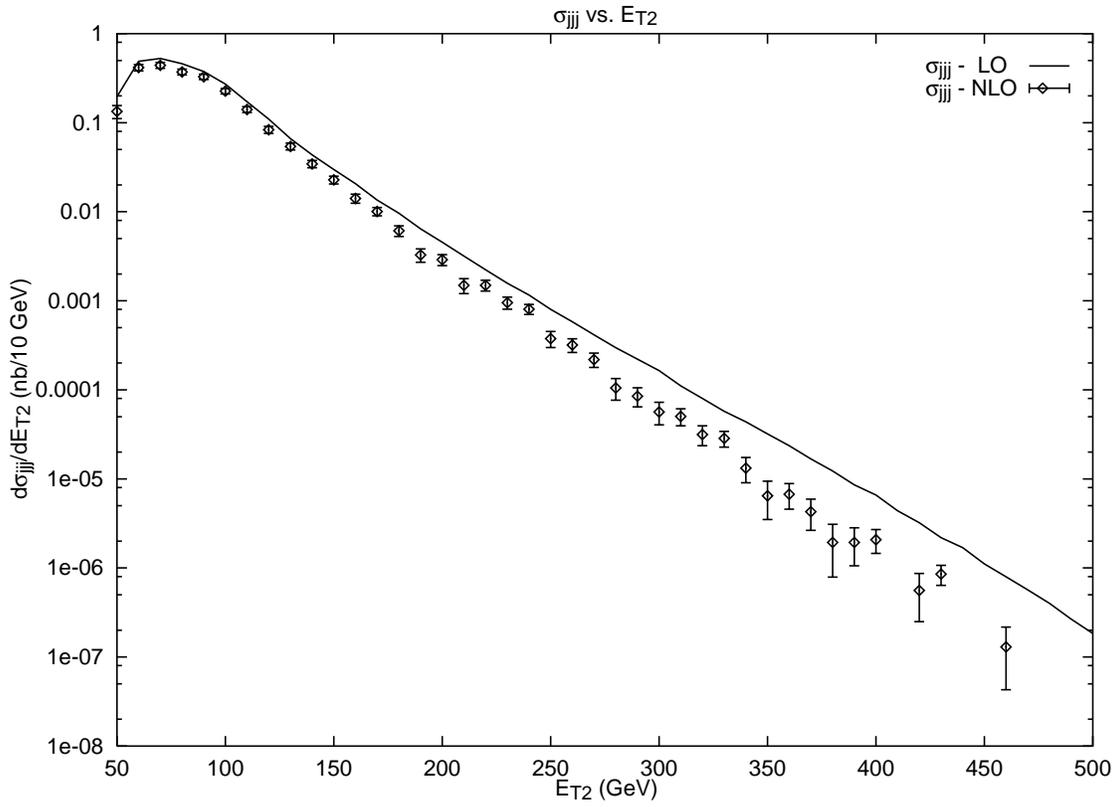}\hfil}   
\caption[]{
\label{fig:ET2plot}
\small Transverse energy spectrum of the second-leading jet.
The leading order result is shown as a solid line.}
\end{figure}

Figure~\ref{fig:ET1plot} shows the transverse energy spectrum of the
leading jet in $E_T$.
There is no indication of any large correction appearing in the jet
spectrum.  The dominant feature of the NLO spectrum is that it is
somewhat softer than the LO spectrum.  That is, NLO predicts that the
spectrum falls more quickly with growing transverse energy than LO.
This is explained in part by the fact that NLO opens up the available
phase space by allowing a fourth jet in the final state.  This same
softening trend is also observed in the transverse energy spectrum of
the second leading jet, shown in figure~\ref{fig:ET2plot}.  Both jet
spectra were computed at $s_{min} = 7.9$
GeV${}^2$.  
\section{Conclusions}
We have successfully built a next-to-leading order event generator for
inclusive three jet production at hadron colliders.  This is the first
NLO calculation of this process to include all parton sub-processes.
Our results indicate that we are correctly canceling the infrared
singularities and therefore obtaining reliable results. With this
calculation we will be able to study many interesting phenomena within
QCD.

\section*{Acknowledgements}
 
W.B.K. wishes to thank L. Dixon and A. Signer for their assistance in
verifying the one-loop matrix elements and Z. Bern for the use of the
UCLA High Energy Physics Group's computing cluster.

This work was supported by the US Department of Energy under grants
DE-AC02-76CH03000 (W.T.G.) and DE-AC02-98CH10886 (W.B.K.).


\begin{references}

%+% 1 ref
\bibitem{Troc}
Z. Tr\'ocs\'anyi, Phys. Rev. Lett. {\bf77}, 2182 (1996) [hep-ph/9610499].

%+% 5 refs
\bibitem{KG}
W.B. Kilgore and W.T. Giele, Phys. Rev. D {\bf55}, 7183 (1997)
[hep-ph/9610433].

%+% 1 ref
\bibitem{kln}
T.~Kinoshita, J.~Math.~Phys. {\bf 3}, 650 (1962);\\
T.D.~Lee and M.~Nauenberg, Phys.~Rev. {\bf 133}, 1549 (1964).

%+% 1 ref
\bibitem{BDK1}
Z.~Bern, L.~Dixon, D.A.~Kosower, Phys.~Rev.~Lett
{\bf 70}, 2677 (1993) [hep-ph/9302280].

%+% 1 ref
\bibitem{BDK2}
Z.~Bern, L.~Dixon and D.A.~Kosower, Nucl.~Phys.
{\bf B437}, 259 (1995) [hep-ph/9409393].

%+% 1 ref
\bibitem{KST}
Z.~Kunszt, A.~Signer and Z.~Tr\'ocs\'anyi, 
{\em Phys.~Lett} {\bf B336}, 529 (1994) [hep-ph/9405386].

\bibitem{BOO}
H. Baer, J. Ohnemus and J.F. Owens, Phys. Rev. D {\bf40}, 2844 (1989);
Phys. Lett {\bf B234}, 127 (1990); Phys. Rev. D {\bf42}, 61 (1990).

\bibitem{KS}
Z. Kunszt and D.E. Soper, Phys. Rev. D {\bf46}, 192 (1992).

%+% 1 ref
\bibitem{GG}
W.T. Giele and E.W.N Glover, Phys. Rev. D {\bf46}, 1980 (1992).

%+% 1 ref
\bibitem{GGK}
W.T. Giele, E.W.N. Glover and D.A. Kosower, Nucl. Phys. {\bf B403}, 633 (1993)
[hep-ph/9302225].

\bibitem{CS}
S. Catani and M.H. Seymour, Phys. Lett. {\bf B378}, 287 (1996)
[hep-ph/9602277];\\ Nucl. Phys. {\bf B485}, 291 (1997);
Erratum-ibid. {\bf B510}, 503 (1997) [hep-ph/9605323].

\bibitem{FKS}
S. Frixione, Z. Kunszt and A. Signer, Nucl. Phys. {\bf B467}, 399
(1996) [hep-ph/9512328].

%+% 1 ref
\bibitem{CTEQ3M}
H.L. Lai, {\it et. al.\/}, Phys. Rev. D {\bf51}, 4763 (1995) [hep-ph/9410404].

%+% 1 ref
\bibitem{EKS}
S.D.~Ellis, Z.~Kunszt and D.E.~Soper, Phys. Rev. Lett. {\bf62}, 726 (1989);\\
S.D.~Ellis, Z.~Kunszt and D.E.~Soper, Phys. Rev. Lett. {\bf69}, 3615
(1992) [hep-ph/9208249].

\end{references}
\end{document}